# A mathematical approach to aspects of LET, micro- and nano-dosimetry in radiation therapy with photons and charged particles

W. Ulmer
MPI of Physics Goettingen Germany

**Abstract**
**Purpose:** The radiation effects induced by $Co^{60}$ serve as a reference system for the consideration of LET and RBE in normal and tumor tissue dose-effect relations are usually handled by the linear-quadratic model (LQ) with the parameters α and β, i.e. $S = \exp(-\alpha \cdot D - \beta \cdot D^2)$. This approximation excellently works up to the shoulder domain. In particle therapy we have strictly differ between RBE in the initial plateau and environment of the Bragg peak. Thus for protons LET and RBE of the initial plateau agree with $Co^{60}$, whereas in the Bragg peak domain both properties are increased,, but RBE of SOBP only varies between 1.1 and 1.17. The RBE of carbon ions is increased once again Their dose-effect curves are much steeper with a rather small shoulder domain due to dense ionizing radiation effect. Thus protons are also dense ionizing in the Bragg peak region, but with rather smaller magnitude compared to carbon ions. A generalization of the LQ-model based on the nonlinear reaction-diffusion model is proposed to describe LET and RBE of dense ionizing particles, which accounts for properties of micro- and nanodosimettry. **Methods:** A linear term of a reaction diffusion formula describes the destroy of cells, the nonlinear term is related to repair and the diffusion term accounts for the density of the radiation damages. **Results:** Based on dose-effect properties of $Co^{60}$ the parameters of dense ionizing particles can be determined and compared with measurement data. **Conclusion:** The local dense of radiation effects and their consequences in RBE and dose effect curves provide a key of understanding modern therapy planning with different modalities and properties of nano-dosimetry are interpreted by mathematical descriptions. The irradiation of spheroids is a feature of micro-dosimetry, whereas intracellular exposure refers to nano-dosimetry.

**key words:** LET, micro- and nano-dosimetry, nonlinear reaction and diffusion, repair, survival functions

## 1. Introduction

As previously shown[1], the LQ-model valid only up to the shoulder can be derived from the model

$$\left. \begin{array}{l} S(D) = \dfrac{A \cdot \exp(-a \cdot D)}{1 + B \cdot \exp(-\rho \cdot D)} \\ \lambda = a + b; \; B = A - 1 \end{array} \right\} \quad (1)$$

S refers to the survival function with S =1 if D = 0 and D to the applied dose. In the high dose region beyond the shoulder eq. (1) assumes the shape of exponential decrease $S = A \cdot \exp(-a \cdot D)$. The LQ-model results from setting[1,2]:

$$\left. \begin{array}{l} \alpha = \lambda / A - b \\ \beta = \lambda^2 \cdot (A-1)/A^2 \end{array} \right\} \quad (2)$$

Thus the condition 1 < A < 2 must hold to fast determine β. Eq. (2) results from an expansion eq. (1) by forming $-\ln S$. Eq. (1) obeys the differential equation;

$$-\frac{dS}{dD} = a \cdot S - \lambda \cdot S^2 \cdot \exp(-\rho \cdot D) \quad (3)$$

It is striking that in the high dose region the nonlinear term $S^2$ is decreasing, and only the first term of eq. (3), i.e. $-dS/dD = a \cdot S \rightarrow S = A \cdot \exp(-a \cdot D)$, is relevant. Detailed applications of eq. (1) to tumor spheroids have previously been given[1,2].

An important feature of eq, (3) is its close relation to nonlinear reaction kinetics, which is commonly applied to cellular regulation processes as well as to problems of biochemistry:

$$\frac{dN}{dt} = \lambda_1 \cdot N - \lambda_2 \cdot N^2 \quad (4)$$

Thus N(t) may refer to either nonlinear kinetics in molecular biology(concentration of biomolecules) or cell biology (growth and dead of cell lines, e.g. spheroids) yielding equilibrium states in both cases. In eq. (4) we have exponential growth (linear term) and decay or death ($N^2$-term) by contact interaction:

$$\left. \begin{array}{l} N(t) = A \cdot \tanh(\beta \cdot t) + B \\ \beta = A \cdot \lambda_2, \; \lambda_1 = 2 \cdot \lambda_2 \cdot B, \; B = A \end{array} \right\} \quad (5)$$

A modification of the condition B = A has been previously stated[2]. With respect to eq. (3) the situation is changed, since the cellular decrease refers to the term - a·S, but due to the factor $\lambda \cdot \exp(-\rho \cdot D)$ the repair part proportional to $S^2$ vanishes with increasing dose D, too. The associated kinetic equation is a modification of eq. (4):

$$-\frac{dN}{dt} = a \cdot N - \lambda \cdot N^2 \cdot \exp(-\lambda \cdot t) \quad (6)$$

This equation can be rescaled to replace the time t by the dose D (if the irradiation time is small compared to the time dependence thereafter (this is not accounted for at this place), and the survival fraction S is given by the ratio $S = N(t)/N_0$ with $N_0 = N$ at $t = 0$. Note: At a small dose rate eq. (6) has to be rescaled in a different way, and a time factor would still appear (low dose irradiation). By taking account for the cellular level (e.g. mono-layers or spheroids) the present state of considerations is restricted to micro-dosimetry, but the intracellular situation requires an extension of eq. (6), namely the role of ionization density by diffusion processes and their connection to the parameters A,B, a and b.

## 2. Methods

### 2.1. Nonlinear repair function with diffusion

The past century has gained the conviction that with regard to regulatory processes and morphologic aspects in molecular biology to pure kinetics such as in eq. (4) and to sole diffusion are insufficient tools to include both for the descriptions of reactions and transport phenomena. Thus the diffusion equation in one space dimension is given by:

$$-\frac{\partial N}{\partial t} + \alpha_0 \cdot \frac{\partial^2}{\partial z^2} N = 0 \quad (7)$$

A well-known solution of this equation is obtained by ($N_0(t)$ is a normalization factor to become a δ-function at $t = 0$):

$$N(x,t) = N_0(t) \cdot \exp\left(-\frac{z^2}{4\alpha_0 t}\right) \quad (7a)$$

However, the complete solution spectrum of eq. (7) with inclusion of a linear kinetic term has previously been given[3]. The simplest generalization of eq. (7) reads:

$$-\frac{\partial N}{\partial t} + \alpha_0 \cdot \frac{\partial^2}{\partial z^2} N = \lambda_1 \cdot N - \lambda_2 \cdot N^2 \quad (8)$$

Thus we shall have to return to the properties of eq. (8) in the present analysis. Eq. (8) has been subjected to some interesting modifications, e.g. the Brusselator[4], and a special type of it has already been rather early by Turing[5] to describe nonlinear reaction-diffusion problems.

With regard to a generalization of eq. (3) we consider the following equation:

$$-\frac{\partial N}{\partial t} + \alpha_0 \cdot g(t) \cdot \frac{\partial^2}{\partial z^2} N = a \cdot N - \lambda \cdot N^2 \cdot \exp(-\lambda \cdot t) \quad (9)$$

The factor function g(t) is included to separate the time behavior in the diffusion term:

$$g(t) = A \cdot \exp(-\lambda \cdot t)/[1 + B \cdot \exp(-\lambda \cdot t)] \quad (9a)$$

Diffusion processes are usually time-dependent, but the use of g(t) takes account of this behavior, and the space-like aspect can be treated independently. However, this choice of g(t) may be a simplifying restriction, but it ensures that all diffusion processes will come to an end at t → ∞ (this fact is also true for the pure diffusion problem according to eq. (7a)) and a connection to eq. (3) can be created. With the aid of g(t) according to eq. (9a) we are able to start with the '*ansatz*':

$$N(z,t) = A \cdot \exp(-a \cdot t) \cdot \frac{1}{1+B \cdot \exp(-\lambda \cdot t)} \cdot H(z) \quad (10)$$

By that, the following differential equation eq. (11) must be fulfilled, which is nonlinear with regard to the space variable x and H''(z) is the second derivative:

$$\left.\begin{array}{l} -B \cdot \lambda \cdot H + \alpha_0 \cdot H'' \\ = -\lambda \cdot H^2; \quad \lambda = a + b \end{array}\right\} (11)$$

We are interested in a particular solution of eq. (11), namely:

$$H(z) = c_z \cdot \operatorname{sech}^2(\alpha \cdot z) \quad (12)$$

The solution function of eq. (12) yields the following conditions:

$$c_z = 6 \cdot \alpha_0 \cdot \alpha^2 / \lambda; \quad B = 4 \cdot \alpha_0 \cdot \alpha^2 \cdot A / \lambda \quad (13)$$

Thus we obtain B < 0. Since the denominator $(1+B)^{-1}$ must be positive, the restriction B > -1 must hold. A is determined by the normalization condition N = 1 (if z = 0 and t = 0). By that, the normalization to determine A at t = 0 and x = 0 can readily be satisfied.

The restriction to the space coordinate x is not required. In the 3D case we have to replace H'' by the Laplace operator Δ, and the argument α·z in eq. (12) is subjected to the substitution::

$$\operatorname{sech}^2(\alpha \cdot z) \to \operatorname{sech}^2(\vec{\alpha} \cdot \vec{z}) \quad (14)$$

The above substitution implies $\alpha^2 = \alpha_x^2 + \alpha_y^2 + \alpha_z^2$ in eq. (13). The general solution of H is presented in an appendix.

## 2.2. Transition to nonlinear survival function with diffusion

In order to receive survival functions S in dependence of the applied dose D, we have to rescale the time variable t appearing in eq. (10) and all solution parameters connected with this equation. This procedure corresponds to a transition from a pure nonlinear kinetic equation with inclusion of spatial diffusion to a generalized survival function S depending on the irradiated dose D. By that, eq. (3) is extended by diffusions related to spatial physical processes recorded by appropriate measurement systems. As already mentioned the survival fraction S is defined by the ratio of the actual cell number N after dose application and initial value $N_0$, i.e. $S = N/N_0$. Then via rescaling we have to account for:

$$\partial S / \partial t = (\partial S / \partial D) \cdot (\partial D / \partial t) = (\partial S / \partial D) \cdot \dot{D} \qquad (15)$$

Thus $\partial D/\partial t$ (or dotted D) refers to the dose rate, which is usually rather high and implies a short irradiation time (the low dose rate irradiation is not considered here). The terms

exp(-a·t), exp(-b·t) and exp(-λ·t) have to be rescaled in the same fashion, too:

$$a \cdot t = a \cdot D / \dot{D} \,;\, b \cdot t = b \cdot D / \dot{D} \,;\, \lambda \cdot t = \lambda \cdot D / \dot{D} \qquad (16)$$

Now the parameters, a, b and λ assume the meaning of a reciprocal dose in the substituted version of eqs. (9, 9a, 10):

$$-\frac{\partial S}{\partial D} + \alpha_0 \cdot g(D) \cdot \Delta S = a \cdot S - \lambda \cdot S^2 \cdot \exp(-\lambda \cdot D) \qquad (17)$$

$$g(D) = A \cdot \exp(-\lambda \cdot D) / [1 + B \cdot \exp(-\lambda \cdot D)] \qquad (18)$$

$$S(\vec{z}, D) = A \cdot \exp(-a \cdot D) \cdot \frac{1}{1 + B \cdot \exp(-\lambda \cdot D)} \cdot H(\vec{z}) \qquad (19)$$

The determination of H(z) or its 3D-extension is identical as already presented.

It is desired that g(D) vanishes for D →∞. A common feature of 1D well as of the 3D case is the argument of the set of sech-functions. For the reason of reduced writing we may reduce us to the 1D case..Thus large α-values imply narrow profiles of the sech-functions to reach the order of $e^{-1}$ by sech(α·z), and the ionization density turns out to be very high, whereas for low α-values the converse is true, and the profiles are significantly broadened. This property is valid for all powers of the form *sech$^n$(α·z).* By that, we may associate increasing α-values for protons or heavy carbons in the environment of the Bragg peak. On the other hand, low α-values implying broad profiles are a characteristic feature of easy ionizing radiation bundles associated with γ-rays of $Co^{60}$ or bremsstrahlung of accelerators. Usually the effectiveness of $Co^{60}$ serve as a reference standard for all other irradiation modalities-.

Note: the dimension of $\alpha_0$ in eqs. (7a) and (9) is (length)/time, whereas in eq. (17) it amounts to (length)$^2$/dose, and α in eqs. (9) and (21) strictly has the dimension 1/length. Thus the behavior of sole diffusion (eq. (7a) ) yields a narrow profile of the concentration N for small $\alpha_0$ value, since it appears in the denominator of this equation. This is the principal difference to the parameter α, which behaves conversely.

We are able to summarize the consequences: The parameters a and b are responsible for the shoulder of the survival function and its steepness at very high doses due to the connection with $\alpha^2$. However, the parameters A and B play the essential role with regard to the normalization, since the inclusion of the diffusion in the nonlinear model affects the normalization, too. The increased complexity can be shown by considering the normalization according to eq. (1), which provides S = 1 for A/(1+B) with B = A-1 and D = 0. By inclusion of diffusion we obtain:

$$S = 1 \rightarrow A \cdot H(0)/(1+B) = 1$$
$$\text{with } D = 0 \text{ and } z = 0 \qquad (20)$$

H(0) is rather easy to evaluate in eq. (12), and eq. (20) yields:

$$S = 1 \rightarrow 6 \cdot A \cdot \alpha_0 \cdot \alpha^2 /(\lambda \cdot (1 + 4 \cdot \alpha_0 \cdot \alpha^2/\lambda)) = 1 \qquad (21)$$

## 2.3. LET of an idealized proton pencil beam

With regard to the LET problem we are able to use some previously obtained results[8], namely the propagation kernel K of the energy transfer from proton to environmental electrons. This kernel has a quantum theoretical background, resulting from the action of the energy exchange operator on to plane waves ψ:

$$\psi = \tfrac{1}{\sqrt{2\pi}} \cdot \exp(i \cdot k \cdot z) \qquad (22)$$

$$\exp(-H/E_{exchange}) \cdot \psi = \exp(-0.25 \cdot s^2 \cdot k^2) \cdot \psi \qquad (22a)$$

$$H = -\tfrac{\hbar^2}{2m} \cdot (d^2/dz^2); \quad s^2 = 2 \cdot \hbar^2 /(m \cdot E_{exchange}) \qquad (22b)$$

The kernel K results from the integration over k according to the spectral theorem and yields:

$$K(s, u-z) = N \exp(-(u-z)^2/s^2);$$
$$N = \tfrac{1}{s \cdot \sqrt{\pi}} \qquad (23)$$

The principal problem with regard to s and the energy $E_{exchange}$ is that **s** depends on the actual energy E of the proton and cannot be a constant value. Therefore we have determined s by a subtraction method of a chain of water molecules starting with a transfer energy of 1 keV at $E_0$. The result of this subtraction method is presented in Figure 1 (it is assumed that the molecules are connected via H bonds and the mean diameter amounts to ca. 0.3 nm). The transfer energy at the end track with z = $R_c$ (CSDA-range) amounts to 30 eV. Thus the resulting energy transfer **$E_{transfer}$** is about 99.97 keV/μm and is in a good agreement with the literature value of 100 keV/μm[7]. This value is also rather constant from 0.0001 cm to 0.001 cm. Thus for brevity we have restricted our considerations on LET to a transfer length of $10^{-3}$ cm and an initial proton energy of 300 MeV at entrance of a phantom.

The LET value of 99.97 keV/μm at end of the proton track is closely connected to the CSDA-approach. In Figures 1 and 1a we show the 'local energy per water molecule ', which results from the energy $E_0$ divided by the number of water molecules per unit length,, obtained by a subtraction method (starting with the lowest energy $E_0$ = 1 keV). extension of $H_2O$: 0.3011 nm, The average diameter of the isolated $H_2O$-molecule amounts to 0.29 nm, but in a chain of molecules connected via H bond the distance is little increased. Thus the CSDA-approach has the advantage that we can start with an arbitrary energy or its related CSDA-value for the length. The result is presented in Figures 1 and 1a.

In order to use analytical methods we pass to either $E_{transfer}$ as a function of the position z and $R_c$ or to he energy E and $E_0$ (initial energy).

$$E_{transfer} = \sum_{k=1}^{3} A_k \cdot \exp(-m_k \cdot (R_c - z)) \; [+A_4 \cdot \exp(-m_4 \cdot (R_c - z))] \qquad (24)$$

The result is represented by eqs. (24, 25) and Tables 1 and 1a.

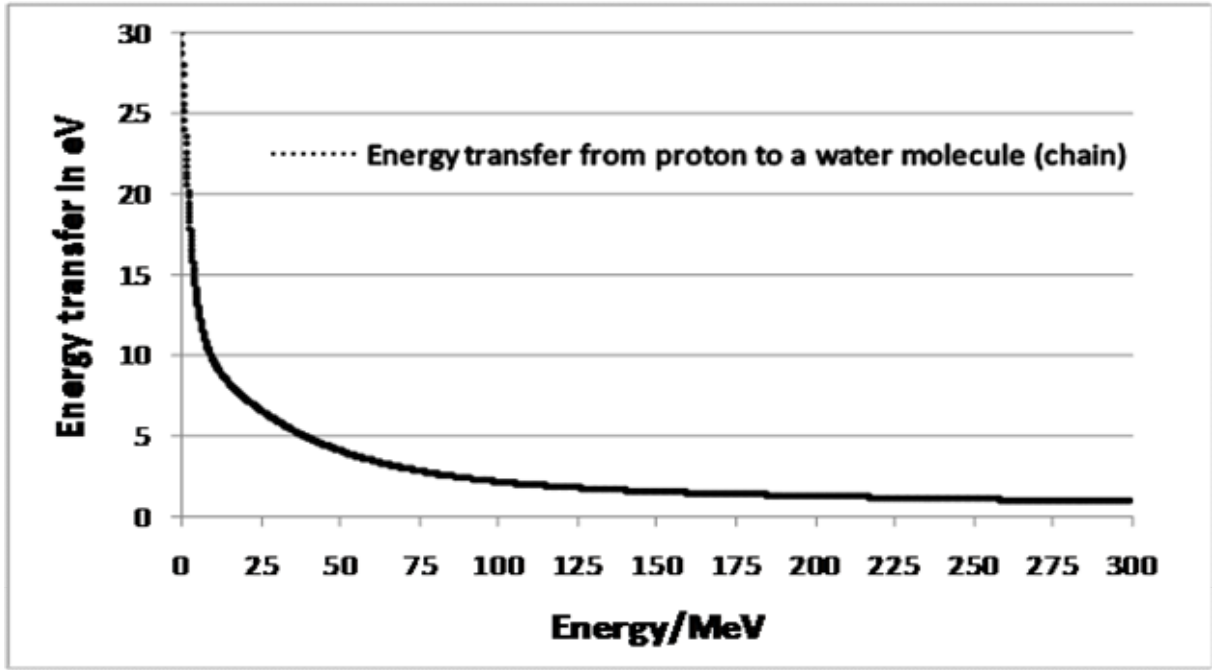

**Figure 1:** The local energy transfer from proton to a chain of water molecules (CSDA- approach).

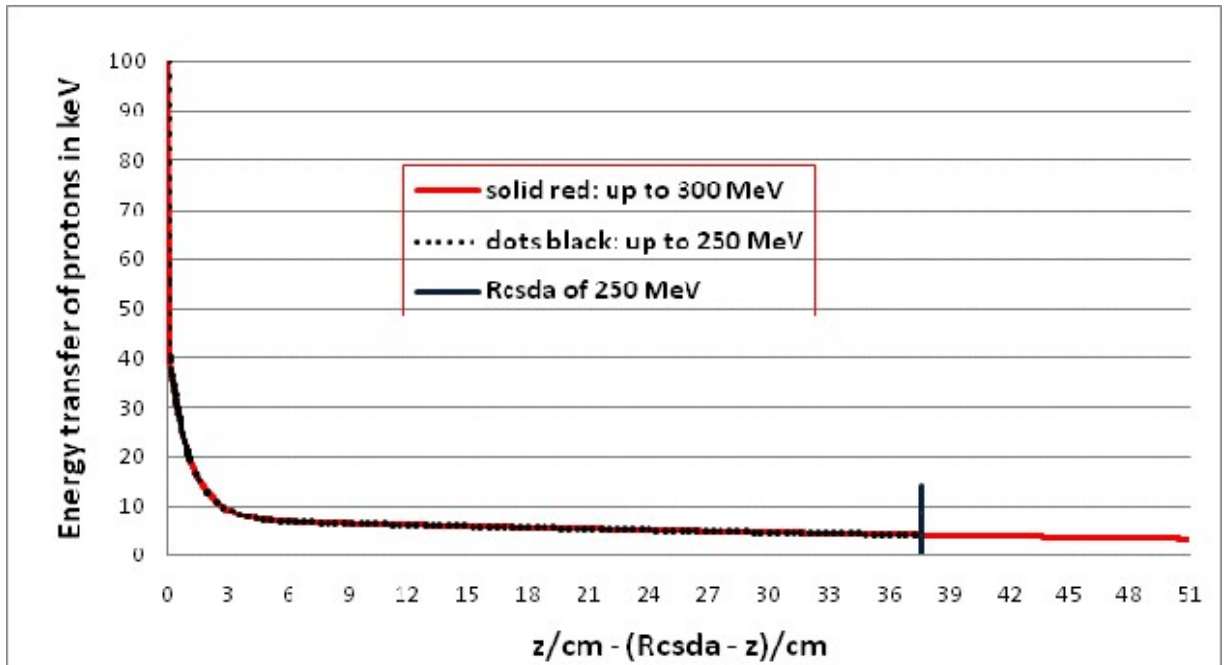

**Figure 1a:** Contents of Figure 1 as a function of z (or $R_c$ - z) , the LET scale is now fixed to $10^{-3}$ cm

**Table 1: Parameters of eqs. (24, 25) valid up to 300 MeV**

| $m_1$/cm | $m_2$/cm | $m_3$/cm | $A_1$/keV | $A_2$/keV | $A_3$/keV |
|---|---|---|---|---|---|
| 40.8 | 0.01473 | 0.88835 | 61.3924 | 7.268 | 31.3396 |
| $E_1$/MeV | $E_2$/MeV | $E_3$/MeV | $A_1$/keV | $A_2$/keV | $A_3$/keV |
| 2.6867 | 330.575 | 33.9245 | 61.3924 | 7.268 | 31.3396 |

**Table 1a: Parameters of eqs. (24, 25) valid up to 600 MeV**

| $m_1$/cm | $m_2$/cm | $m_3$/cm | $A_1$/keV | $A_2$/keV | $A_3$/keV |
|---|---|---|---|---|---|
| 45.8 | 0.03664 | 1.0155 | 61.37 | 5.588 | 30.3226 |
| $E_1$/MeV | $E_2$/MeV | $E_3$/MeV | $A_1$/keV | $A_2$/keV | $A_3$/keV |
| 2.6147 | 206.17 | 31.7635 | 61.37 | 5.588 | 30.3226 |
| $m_4$/cm: | 232588.0 | $A_4$/keV→ | 2.7194 | $E_4$/MeV→ | 1872.68 |

Please note that by taking account of $A_k$ and $m_4$ modifications of the remaining terms are involved (Table 1a). It is possible to represent $E_{transfer}$ as a function of the actual energy E. Then we have to replace $R_c$ by the initial energy $E_0$, z by the actual energy E and $1/m_k$ by the corresponding energy parameters $E_1$, $E_2$, $E_3$ (and $E_4$) of the energy ranges.. The resulting modification of eq. (24) reads:

$$E_{transfer} = \sum_{k=1}^{3} A_k \cdot \exp(-(E_0 - E)/E_k) [+A_4 \cdot \exp(-(E - E_0)/E_4)] \quad (25)$$

It may seem that $E_{transfer}$ from proton to environmental electrons shows mono-energetic properties. This is not true, the energy dependence of $E_{max}$ is given according to eq. (26):

$$E_{max} = s_1 \cdot E + s_2 \cdot E^2 + s_3 \cdot E^3 + s_4 \cdot E^4 \quad (26)$$

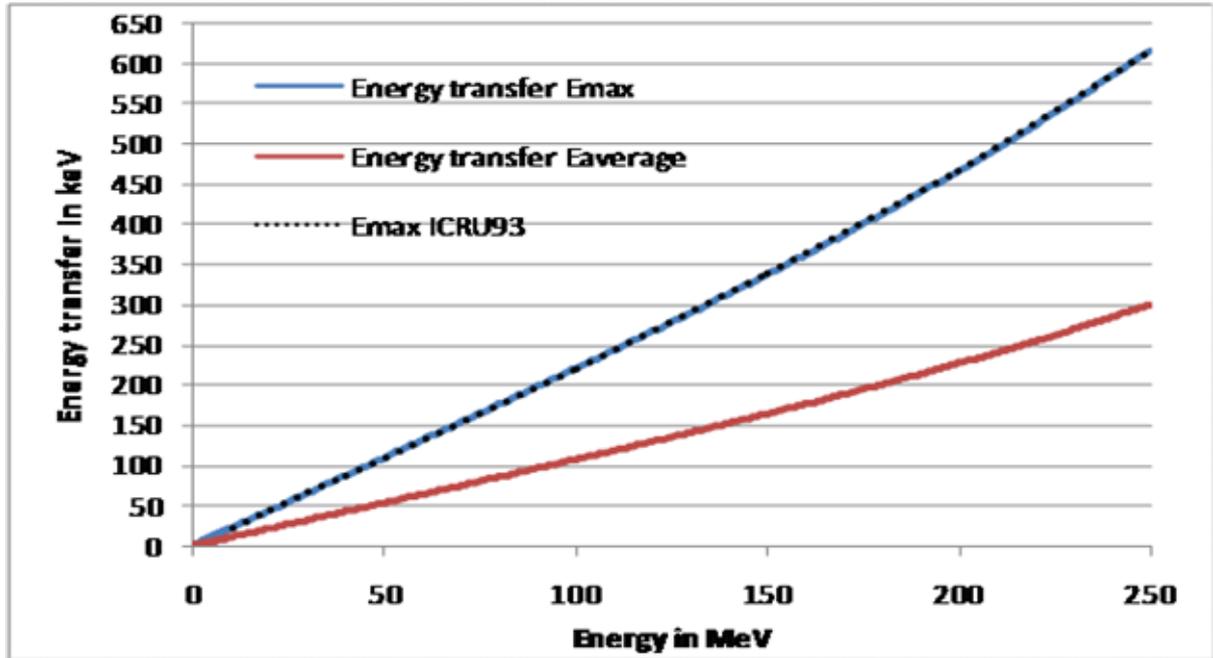

**Figure 2:** Maximum energy $E_{max}$ from proton to environmental electrons (blue curve, dots: ICRU93 and a average electron energy $E_{Average}$ (red curve and eq. (27)).

The parameters of eq. (26) are: $s_1$ = 2.176519870758; $s_2$ = 0.0049990000698; $s_3$ = -0.00000004502; $s_4$ = 0.000000017988: they result from a fit of a numerical adaption of an analytical integration of the Bethe-Bloch equation[8]. However, we need also the average transfer energy $E_{av}$ from protons to environmental electrons; this is performed via the formula:

$$E_{proton} = C \cdot \int_0^{E_{max}} erf(E/E_{max}) dE = C \cdot E_{max} \cdot (erf(1) - (1 - e^{-1})/\sqrt{\pi})$$

$$E_{proton} = E_{av} \cdot C \rightarrow E_{av} = E_{max} \cdot (erf(1) - (1 - e^{-1})/\sqrt{\pi}) =$$
$$0.486064 \cdot E_{max} \quad (27)$$

Eq. (26) results from a theoretical calculation based on an analytical integration of Bethe-Bloch equation and the comparison with ICRU49 data[8,8]. Thus this Figure2 refers to the so-called δ-electrons released by the interaction of protons with the environmental electrons.

In order to pass from keV/length (length = 0.001 cm) to dose in Gray referring to the mass of the volume (length)$^3$ and density 1 (water), we have to carry out a simple calculation and receive the following result: We need **308** protons to obtain 2 Gy at the Bragg peak. If one takes account for the average diameter of a human cell[14], which amounts to $5 \cdot 10^{-3}$ cm, then it would be reasonable to regard a constant proton fluence within the square of $10^{-2} \cdot 10^{-2}$ cm$^2$, which can be realized by **308000** protons of the corresponding dose volume. Thus we would have reached the domain of nanodosimetry, and it is assumed that the suitable energy is obtained by a range shifter.

Figures 3 and 3a provide an interesting result with regard to comparisons of Co$^{60}$ γ-rays and protons. Both Figures result from an analysis of the well-known Klein-Nishina formula. Thus it is usually assumed that the LET of Co$^{60}$ is assumed to be 0.3 keV/μm or 3 keV/0.01 mm, whereas for protons LET in the initial plateau is stated with 3 - 6 keV/0.01 mm. On the other hand, the LET of γ-rays of the order 200 keV - 300 keV is stated as 10 - 20 keV/0.01 mm. Thus we have examined previous GEANT4-Monte Carlo calculations and could verify that the usual assumption for Co$^{60}$ γ-rays is true in the case of Figure 3, whereas in the case of Figure 3a the recoil photons have to be included. They are of the order 200 keV - 300 keV and depend on the reflection angle. Please note that energy-momentum conservation has to be accounted for with regard of the Compton effect for the system 'photon - electron'. Thus Figure 3 considers the case, where the incident photon is not or only slightlydeflected by the interaction with electrons, whereas Figure 3a includes the maximal energy transfer to electrons, and the recoil photon is backscattered with an energy in the keV-domain.

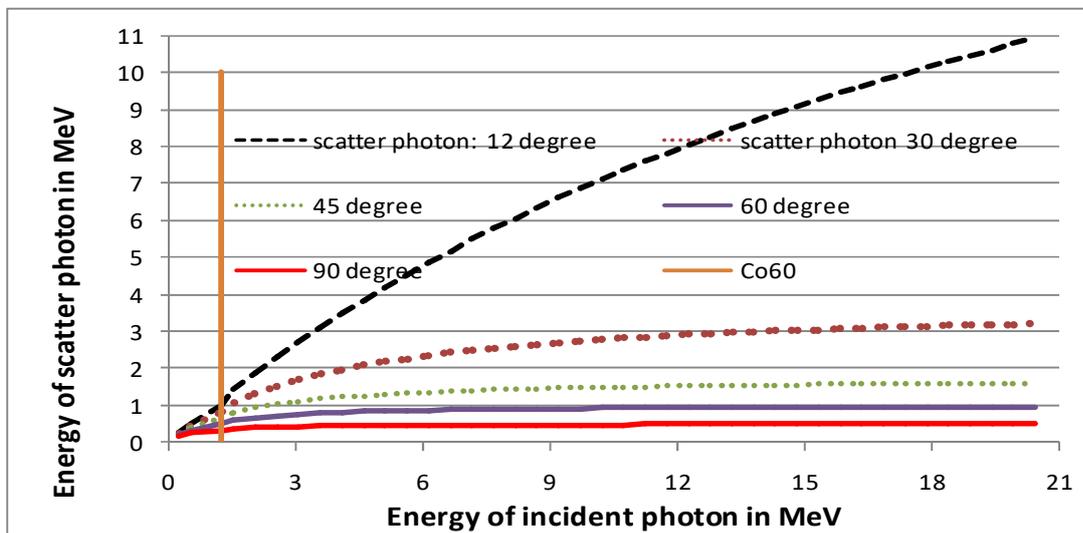

**Figure 3:** Energy of the recoil photons by Compton scatter - angle distribution of forward and lateral scatter.

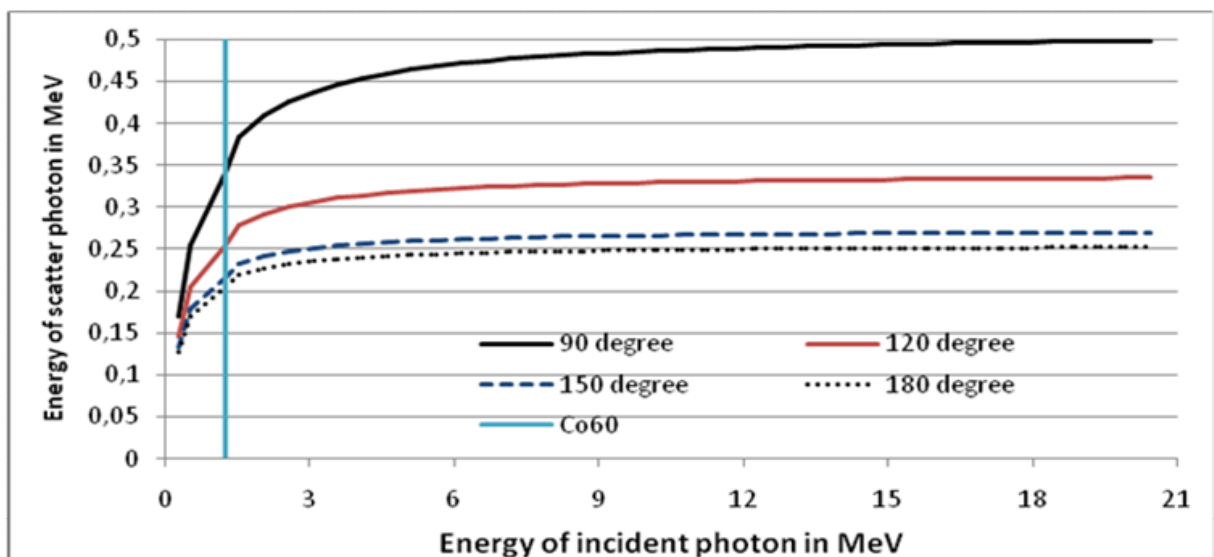

**Figure 3a**: Energy of the recoil photons by Compton scatter - angle distribution of backward and lateral scatter.

# 3. Results

## 3.1. The role of diffusion in micro- and nanodosimetry: dense ionizing radiation and LET

There exists a close relation between LET and dose-effect relationship (RBE) in micro- as well as in nano-dosimetry and a detailed review of this issue is rather instructive[7]. It is convenient to use $Co^{60}$ as a reference standard for LET and RBE in radiobiology/radiotherapy. The LET of $Co^{60}$ γ-rays amounts to 0.3 keV/μm, whereas for protons (10 MeV) in the environment of the Bragg peak and neutrons it amounts to 100 keV/ μm. However, this value is continuously for higher proton energies due to energy straggling, which is also increasing with energy. A further reducing effect results from the scatter influences of the beam-line of protons leading to broadening of the Bragg peak. This can be verified in Figures 4 - 6, where the differences of the Bragg curves are presented.

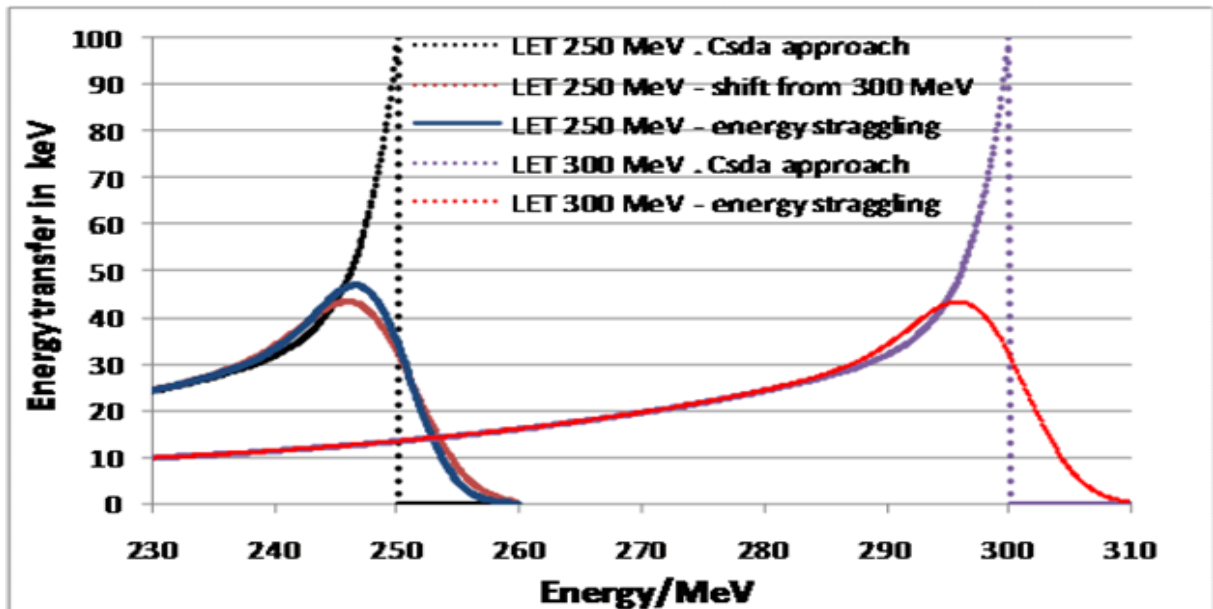

**Figure 4:** 300 MeV protons (CSDA and energy straggling) and range shift to 250 MeV.

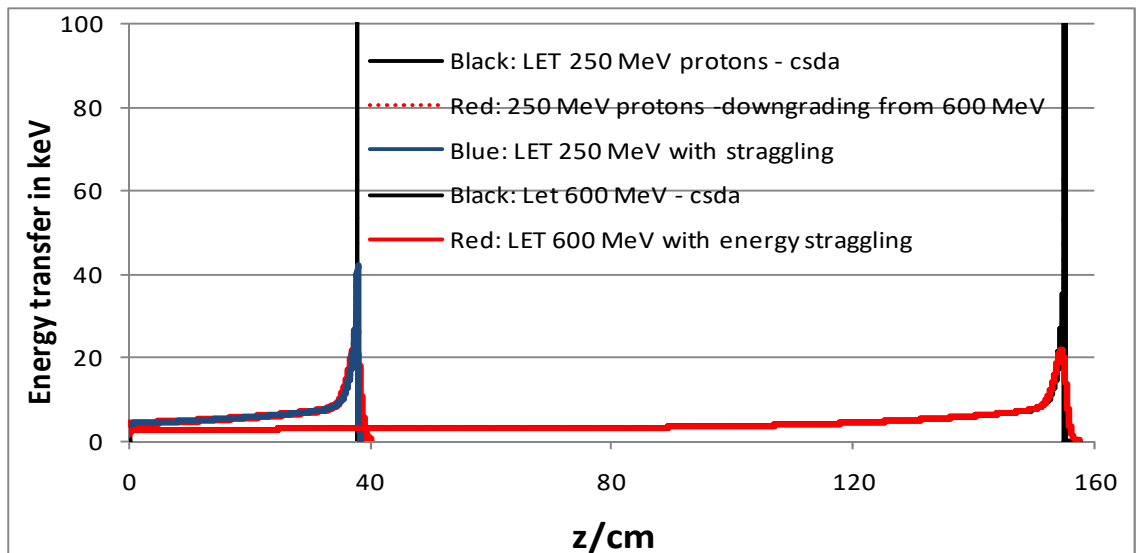

**Figure 5:** Modification of figure 1a and extension to 600 MeV - normalization to one proton 600 MeV protons with energy straggling and without (CSDA) and range shift to 250 MeV via downgrading.

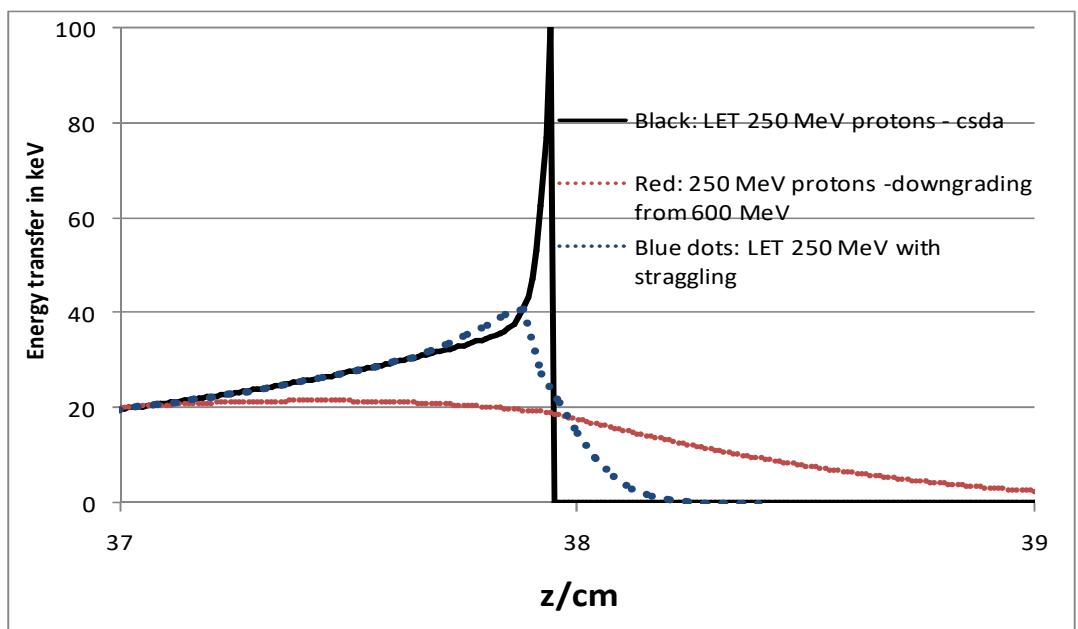

**Figure 5a:** Domain of the Bragg peak of 250 MeV and range shift from 600 MeV.

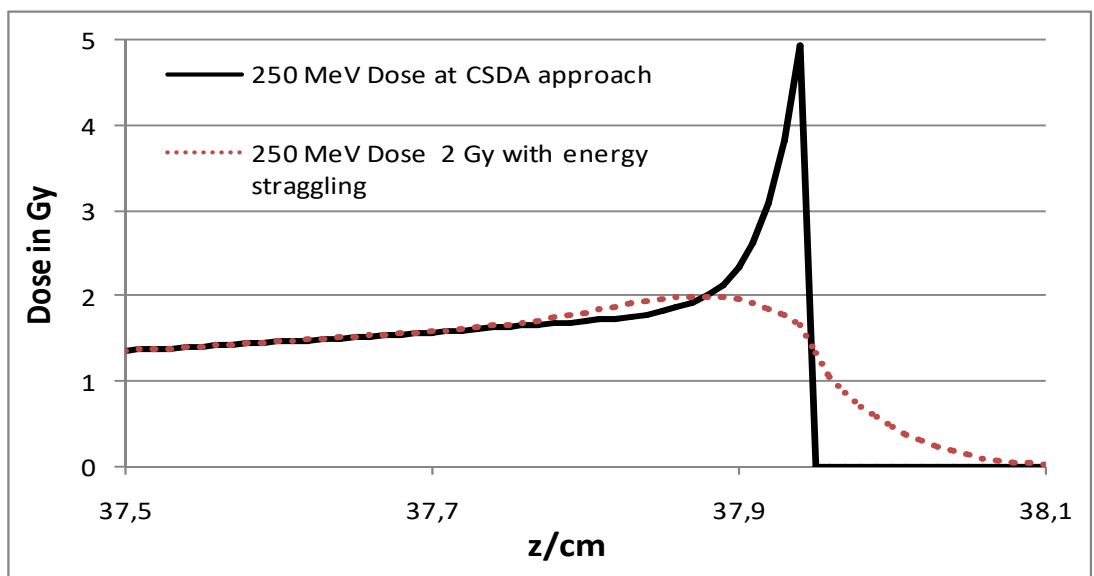

**Figure 5b:** 2 Gy at the Bragg peak (250 MeV) of the CSDA approach (black curve) and curve with energy straggling.

The idealistic case of mono-energetic primary protons, which would provide the highest LET in Bragg peak domain, cannot be reached for the following reasons:
1. The cyclotron itself cannot produce mono-energetic protons; if a synchrotron is used then only the half-width of the Gaussian energy distribution is reduced. 2. The protons originated by cyclotrons have usually to be downgraded by range shifters leading to further broadening of the energy spectrum. 3. The passage of protons through media (water, patients, etc.) is connected with lateral scatter and energy straggling. 4. Last but not least the secondary protons induced by nuclear reactions are a further source broadening of the energy spectrum. All theses influences incorporate noteworthy reasons that the LET of 100 keV/μm is far from realistic conditions given in radiotherapy with protons.

The pristine Bragg curve according to Figure 6 is taken from a previous publication[9] and refers to a 250 MeV cyclotron (Varian-Accel). Thus the measurement data only contain lateral scatter, energy straggling in water and influence of the nozzle. The LET-value in the Bragg peak domain amounts to 72.2 keV/μm. If the desired energy has to be determined by a range shifter via downgrading, then the

LET in the Bragg peak region is additionally decreasing in a significant way. Thus LET values of 50 - 60 keV/µm are realistic. Some pristine Bragg curves resulting from further downgrading can be verified in the previous publications[8, 9].

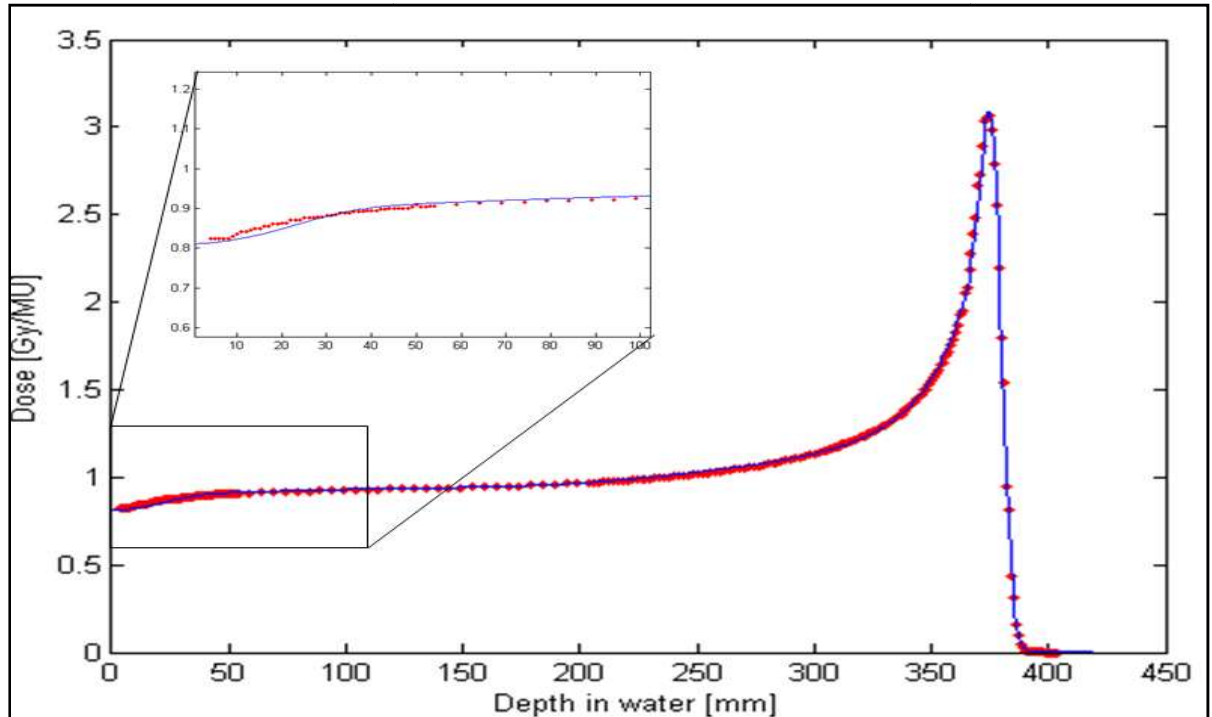

**Figure 6**: Pristine Bragg curve under realistic conditions:. 250 MeV protons from a Varian-Accel cyclotron (calculation solid line and measurements dots).

The consideration of pristine Bragg curves created under realistic conditions shows that an analysis of the LET and, by that, the determination of RBE is not a simple task, because LET significantly depends on the beam-guide. On the other hand, RBE depends on the cell-lines and tissue-specific properties, and it is generally assumed that a correlation between LET and RBE exists see. (e.g. . in WEB[10], where various references can be found). However, the clinical RBE = 1.1 assumed in TOPAS[10] has already led to rejections[11]. However, it should be mentioned that the debate referring to an accurate RBE-value is certainly a complex problem, since in clinical applications usually a SOBP makes sense and this profile consists on linear combinations of different Bragg curves inclusive scatter contributions, where all LET sections play the non-negligible role. However, the principal difficulty results from the transition from cellular assays (e.g. mono-layers, tumor spheroids in micro-dosimetry) to the complex properties of human tissue. Therefore profound clinical studies are only able to lead to more clearness.

The transfer energy of protons in dependence of their actual energy E is characterized by a Landau spectral distribution, and only for actual energies below 100 MeV the spectral distribution tends to a Gaussian shape. A result of this behavior is the buildup effect, which can excellently be verified in Figure 6.

Rather early investigations of RBE of 90 MeV protons have already presenteed[12]. The cell lines '*HeLa cells and Chinese hamster cells* ' have been treated with 0.8 Gy under different conditions (plateau, Bragg peak and distal end). By determining the survival fraction in the plateau region, t he survival in the Bragg peak via colony forming ability was about 1.5 (micro-dosimetry) and the chromosome aberration damage amounted to about 1.8 - 2. However, we should be aware of that all differences are still rather small at a dose D = 0.8 Gy, and the application of 2 Gy should lead to larger differences due to the exponential behavior of the survival function. The dependence of the ATP-concentration of cell cultures on the survival has also been verified many years ago[13], and it should be possible to determine radiation effects induced by protons with the help of MR-spectroscopy. Since the ATP-concentration is an intracellular property, its determination as a function of repair can be placed between micro- and nano-dosimetry.

Since our starting point of proton LET is the CSDA-approach with 100 keV/μm, the relationship to conventional dosimetry is the following factor:

$$E_{tr} = 100 \; keV = 1.602176565 \cdot 10^{-14} \; Joule \quad (28)$$

The transferred energy per length ($L_0 = 1$ μm) from proton to environment (mainly electrons) should finally undergo a transition to specific volume, where the dense ionizing energy is stored, i.e.:

$$\frac{1}{L_0} \cdot \int_0^{L_0} E_{trr} \, dz = \frac{1}{V} \cdot \int_0^V E_{ttrr} \, dV \quad (29)$$

This means that we have to look for the suitable volume, where the energy $E_{transfer}$ is stored, i.e. we pass from energy/length to energy/volume (energy density). According to preceding formulae, we suggest that this connection is mediated via the parameter α and $V^{-1}$ is defined by $α^3$ (if $α_x \neq α_y \neq α_z$ we have to take the **root** $α = (α_x^2 + α_y^2 + α_z^2)^{1/2}$. This implies that $α^3 \cdot E_{tr}$ yields the energy density per volume V.

### 3.2 SOBP and LET

Figure 7 shows the SOBP resulting from a shift of a mono-energy of $E_0 = 250$ MeV. A further downgrading would not affect the SOBP. The considered case would be suitable for SRT with protons.

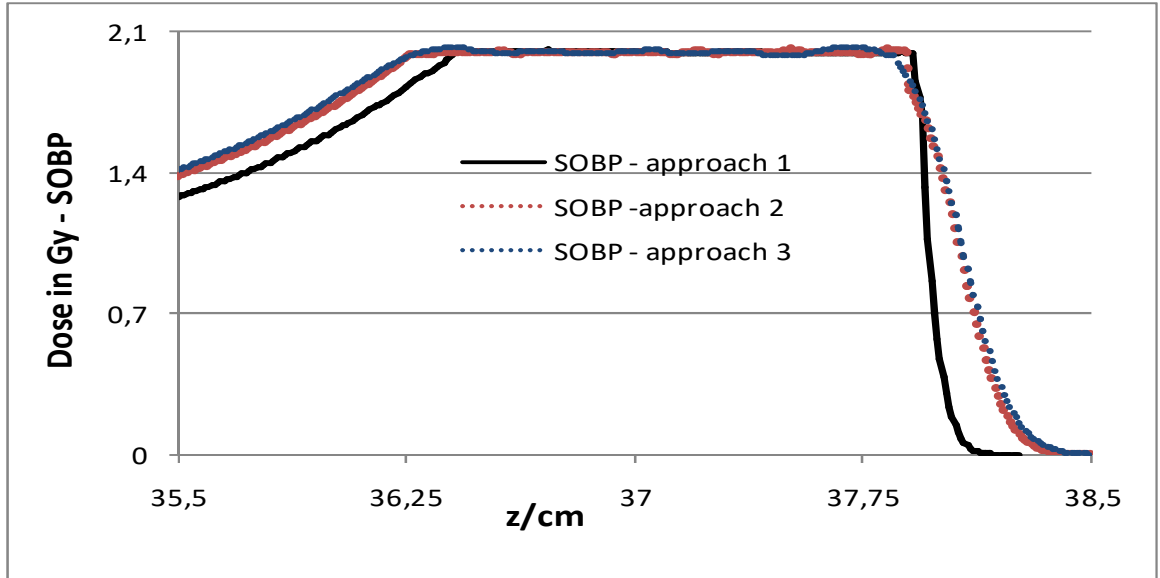

**Figure 7:** SOBP of 2 Gy based on 250 MeV protons calculated by 3 different conditions.

Thus approach 1 uses 15 Bragg curves (length: 1.5 cm) refers to the idealistic situation: the impinging proton beam only contains energy straggling of the water-equivalent shift material, case 2 refers to the realistic properties of Figure 6, where energy straggling of the beam-guide is included.

Approach 2 only uses 6 Bragg curves providing 1.66 cm length. This realistic case would be suitable for proton SRT. Approach 3 is similar like approach 2, but the proton beam is downgraded from 600 MeV.

Thus approach 1 would lead to an averaged LET of 20.26 keV/$10^{-3}$ cm, in the realistic case 2 this value assumes 18,37 keV/$10^{-3}$ cm, The total energy deposition amounts for Figure 7 $E_{st} =$

34.399 MeV. Please note that the total average LET amounts to $LET_{Av}$ = 6.58935 keV/$10^{-3}$ cm.

In literature[7] it is stated that $Co^{60}$ γ-rays are connected with LET = 3 keV/$10^{-3}$ cm . This is, however, restricted to that situation, where the back scattered recoil photons are neglected. Many therapy planning systems neglect this effect, and only forward scattered photons are accounted for. An examination gave that this value should be corrected to assume 3.37 keV/$10^{-3}$ cm.

It is very exciting, that with a permanent energy transfer LET = 3.37 keV/$cm^{-3}$ the total energy of 1.25 MeV would already consumed at 0.234962406 cm, if the energy transfer mechanism of photon electron would be identical with the mechanism of protons, and 6 photons placed with this distance would provide a length of 1.4097736 cm. This is, however, a severe contradiction to reality and, by that, the difficulty of a reduction of the proton RBE to photon RBE emerges! Therefore the solution of the contradictions is given by the Klein-Nishina cross-section formula to determine the number of photons and the probability of LET = 5.32 keV/$10^{-3}$cm per one photon. Assume the target domain for $Co^{60}$ irradiation between z = 5 and 6.5 cm, then the difference of loss of fluence amounts to 0.057356351. Thus we need a fluence of 112 (rounded) photons in an area of $10^{-3} \cdot 10^{-3}$ $cm^2$ to reach a comparable effect as via one proton.

### 3.3. Principal results

On the basis of this result the situation for $Co^{60}$ X-rays and the relationship to protons is quite different. In order to store with a LET of 5.32 keV/$10^{-3}$ cm the total photon energy of 112 photons amounts to 140 MeV instead of 1.25 MeV. If one wants to pass to dose in Gy a much higher number of photons are required - this is similar in the case of protons, but with regard to LET we do not need this information. Then with the probability behavior of photons according to the formula of the fluence Φ we calculate the total energy transfer for this distance with the help of Klein-Nishina formula:

$$\left. \begin{array}{l} \Phi = \Phi_0 \cdot \exp(-\mu \cdot z) \\ \mu = 0.0510825584 \quad cm^{-1} \end{array} \right\} (30)$$

This gives for the calculated photon number an energy of $E_{stored}$ = **8.029896 MeV** for the length of 1.409 cm; the total initial energy amounts to **140 MeV** with regard to the total photon fluence.

The general formula according to eqs. (20, 21) reads:

$$2 \cdot A \cdot \alpha_0 \cdot \alpha_z^2 / \lambda = 1 \quad (31)$$

In the following we assume that the parameter $\alpha_z$ is determined by the ratio LET/stored energy in the target domain (LET/$E_{st}$).. The dimension of $\alpha_0$ is $length^2$ dose, whereas $\alpha_z^2$ has the dimension 1/$length^2$. Since the reference system is $Co^{60}$ should provide $\lambda = \lambda_0$, and LET is given by the transferred energy $E_{tr}$/length, i.e., $E_{tr}/l_c$ ($l_c = 10^{-3}$ cm), the following fixation is useful, so far $\lambda \neq \lambda_0$:

$$\left. \begin{array}{l} \alpha_z = C \cdot l_c^{-1} \cdot \dfrac{E_{tr,0}}{E_{st,0}} \cdot \dfrac{E_{st}}{E_{tr}} \\ \\ C = \lambda / \lambda_0 \end{array} \right\} (31a)$$

Thus we put the further fixation concerning the parameter *A*, we assume that A of protons is identical to the reference A of $Co^{60}$, $\alpha_0$ is defined by:

$$\alpha_0 = \tfrac{1}{2} \cdot l_c^2 \cdot \lambda_0 / A \quad (31b)$$

Then by setting $\lambda = \lambda_0$ and $E_{st} = E_{st,0}$, $E_{tr} = E_{tr,0}$ all parameters of the reference system are given via eqs. (31 - 31b).

Based on Figure 7 (if reduced to 1.66 cm) the following numerical values are valid:

$$\left. \begin{array}{l} Co^{60} : E_{tr,0} = 3.347 \; keV \; ; \; E_{st,0} = 5087.002 \; keV \\ \\ protons : E_{tr} = 18.37 \; keV / 10^{-3} \; cm \; ; \; E_{st} = 31979.003 \; keV \end{array} \right\} \quad (32)$$

In approach 3 the proton parameters are slightly modified: $E_{st}$: 30655.012 keV and LET: 17.95 keV/$10^{-3}$ cm. Using these numerical values the λ-value of protons have to be scaled by $\lambda = 1.33 \cdot \lambda_0 = 1{,}33 \cdot (a_0 + b_0)$ in the case of approach 2, whereas approach 3 yields λ = 1.28.

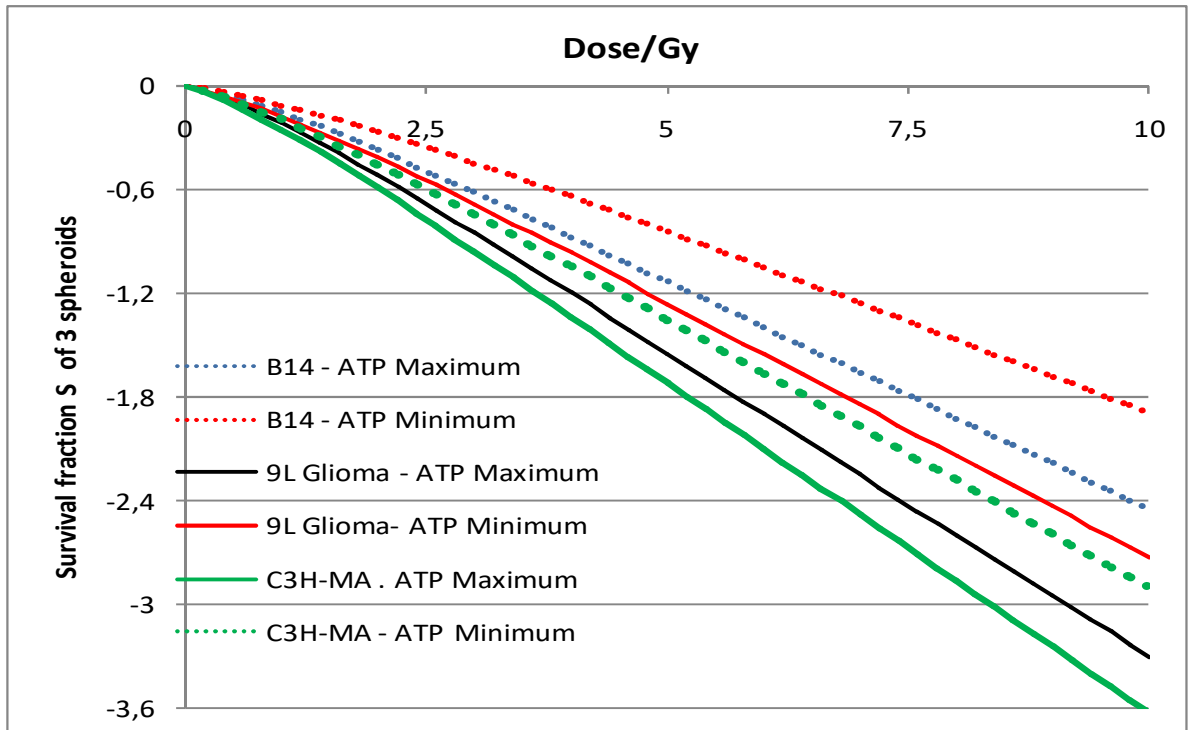

**Figure 8:** Survival fraction of the spheroids B14, 9L Glioma, C3H-MA with maximum and minimum ATP concentration.

We now apply these proton results to 3 previously analyzed[1,2] tumor spheroids [B14, 9L Glioma, C3H-MA). Figure 8 clearly shows the uncertainty of the determination of a survival fraction S. Based on 31P-NMR spectroscopy the ATP consumption has been determined before irradiation, and a significant difference between the maximum of the ATP concentration and the related minimum is a striking feature. It is also possible that further influences play a role with regard to the survival fraction S. Therefore we have accounted for the mean values between the two extreme cases to determine the behavior under proton irradiation (Figure 9).

The RBE of the 3 spheroids is given in Figure 10. This Figure provides a striking information with regard to eq. (1). Thus for doses > 6.5 Gy only the contribution $A \cdot exp(-a \cdot D)$ becomes significant, and the nonlinear part referring to repair and repopulation is vanishing. This fact indicates that we have passed from micro- to nano-dosimetry, where only inner cellular damages are accounted for. The difference between $Co^{60}$ and protons tends to RBE = 1.4, whereas for low doses this factor is more than 1.8. The calculation parameters for the 3 spheroids are presented in Table 2.

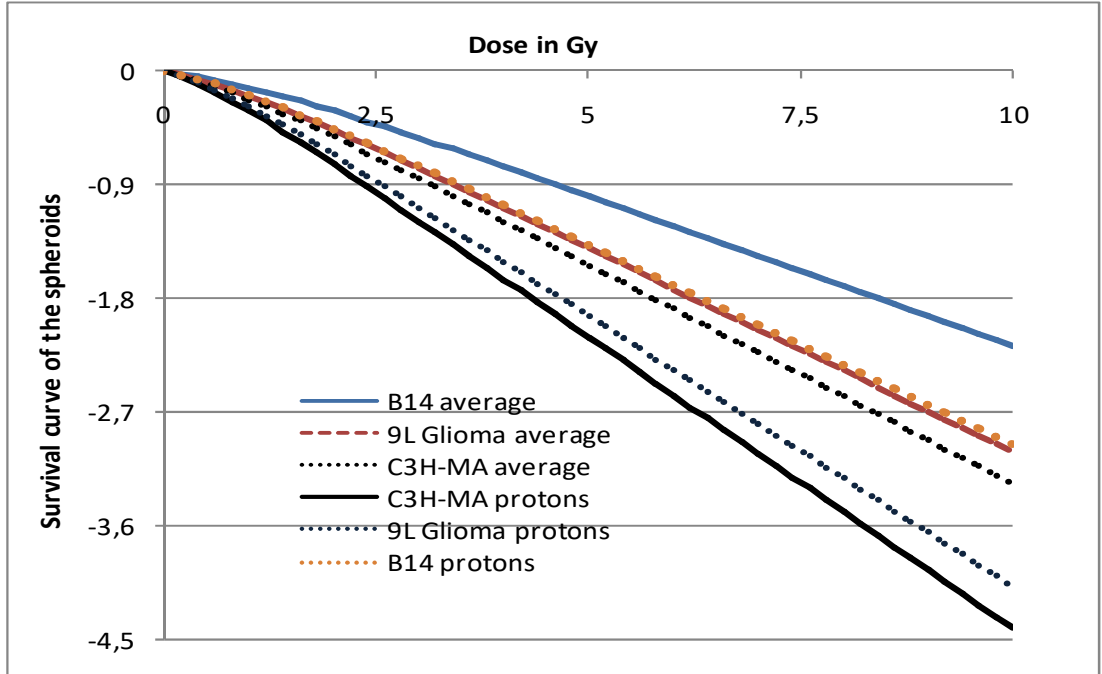

**Figure 9: Survival** curves of the mean values of the 3 spheroids and the related proton curves

**Table 2: Parameters of the spheroids (ATP maximum, ATP minimum and ATP average and HeLa cells).**

| B14 | A | a | b | | A | a | b |
|---|---|---|---|---|---|---|---|
| max | 1.51 | 0.606 | 0.394 | min | 1.66 | 0.487 | 0.198 |
| B14 av | 1.585 | 0.546 | 0.296 | | | | |
| 9L -Gli | | | | | | | |
| max | 1.59 | 0.8668 | 0.372 | min | 1.63 | 0.677 | 0.224 |
| 9L-Gli av | 1.61 | 0.772 | 0.298 | | | | |
| C3H-MA | | | | | | | |
| max | 1.54 | 0.879 | 0.388 | min | 1.59 | 0.716 | 0.212 |
| C3H-MA av | 1.565 | 0.797 | 0.3 | | | | |
| HeLa | 1.07 | 0.47658 | 0.229 | | | | |

Figures 11 principally refers to cervical uteri tissue (based on HeLa cells[16,17,18]) irradiated with protons and the corresponding survival curve (Figure 12), but this Figure also provides the comparison with the cell culture of HeLa according to Figure 7. The parameters are stated in Table 3. A striking feature is the transition from cell culture to clinical extension of the target. The RBE (2 Gy) is reduced from 1.33 (approach 2) to 1.19 and from 1.28 (approach 3) to 1.155. Thus the difference between the 2 cases is rather of minor order and explains the rather small increase of RBE by protons compared to $Co^{60}$. A theoretical access to irradiation parameters of cervix uteri has previously presented[19], which provided a first indication of the corresponding order. If we determine from the parameters a, b, and A the parameters of the LQ-model, we obtain $\alpha = 0.573$ $Gy^{-1}$ and $\beta = 0.066$ $Gy^{-2}$ in approach 2 and $\alpha = 0.551$ $Gy^{-1}$ and $\beta = 0.063$ $Gy^{-2}$ in approach 3. A literature value[16] is stated by $\alpha = 0.53 \pm 0.12$ $Gy^{-1}$ and $\beta = 0.084 \pm 0.025$ $Gy^{-2}$. With regard to $Co^{60}$ we have used the following bases: $\alpha = 0.43042 \pm 0.1$ $Gy^{-1}$ and $\beta = 0.0404 \pm 0.01$ $Gy^{-2}$. With help of the bars the reference values[19] are confirmed.

It should be mentioned that in the 3D case the relation (31b) reads:

$$l_c^2 = (l_{c,x}^2 + l_{c,y}^2 + l_{c,z}^2) \quad (33)$$

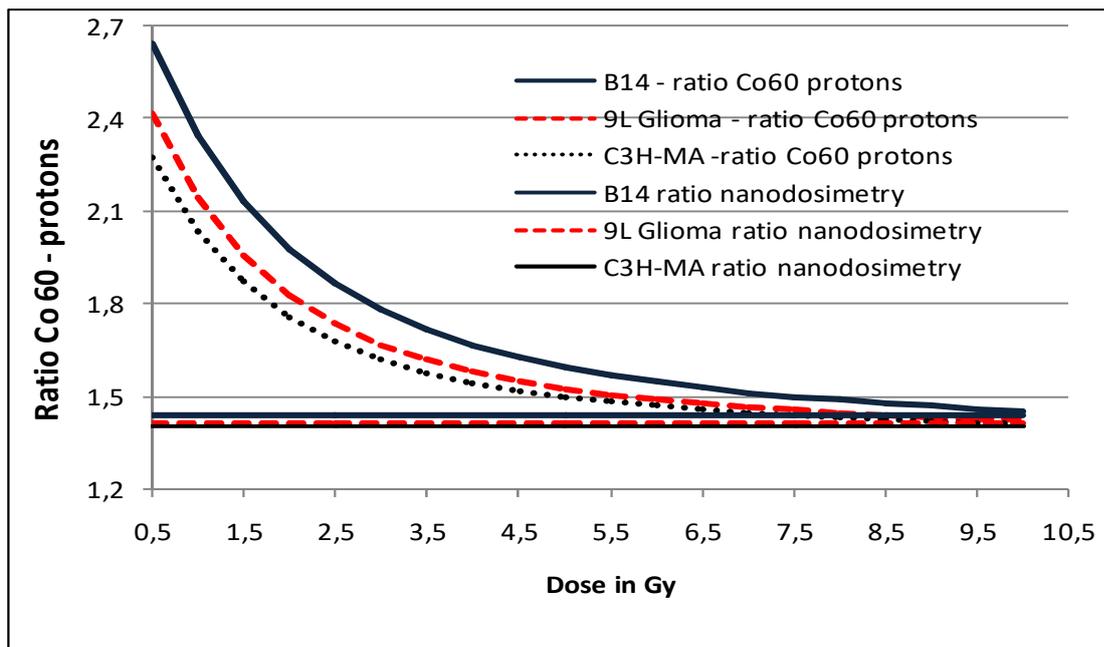

**Figure 10:** RBE of protons (micro-dosimetry) of the 3 spheroids relative to $Co^{60}$ (micro-dosimetry).and nano-dosimetry). The nano-dosimetry results from $S = A \cdot \exp(-a \cdot D)$; the ratio remains constant.

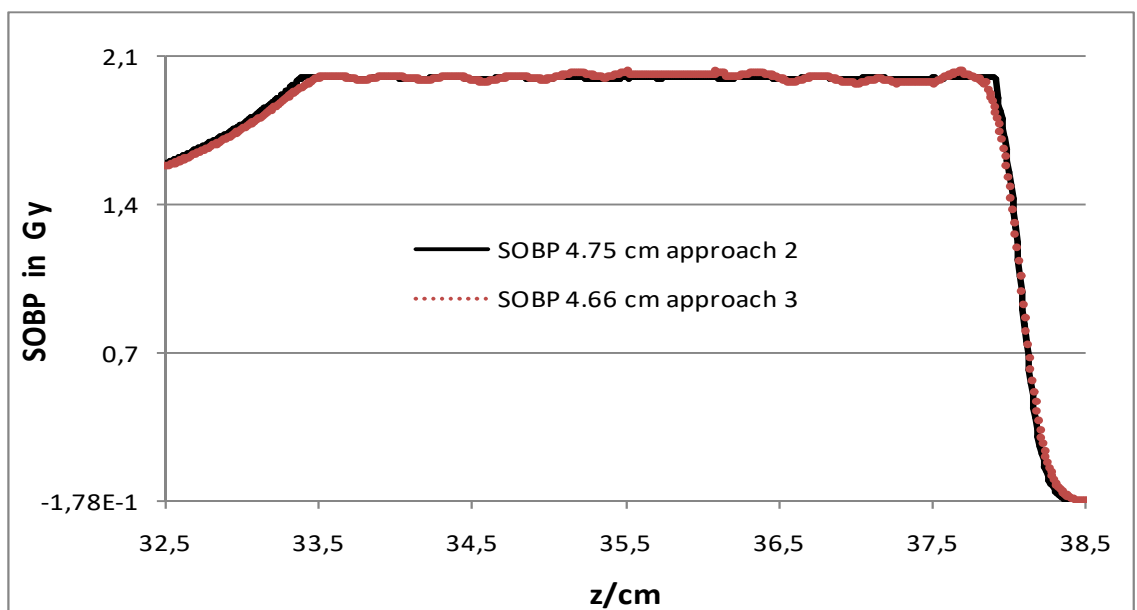

**Figure 11:** SOBP according Figure 7 of a proton pencil beam with acceptable clinical conditions.

However, so far we are able to restrict our analysis to parallel beams, the restriction to the z-direction makes sense.

With regard to protons we have to be aware of that the target length and volume might be decisive. The stored energy $E_{st}$ will increase, and the value for proton LET will decrease in the corresponding manner. The parameters $\alpha_x$ and $\alpha_y$ mainly determine properties in 3 dimensions of the radiation volume inclusive the penumbra, but the principal aspects of the connection between LET and radiobiological parameters remain unchanged.

Please also note: The corresponding actual value is rigorously depending on the beam guide. This fact explains the comparably small increase of proton RBE of 1.1 - 1.17 compared to $Co^{60}$. The most worse case results from downgrading of a much higher initial proton energy, e.g. 600 MeV. In a further communication we shall consider dose-effect relations and RBE, when passing from $Co^{60}$ or linac to proton radiotherapy.

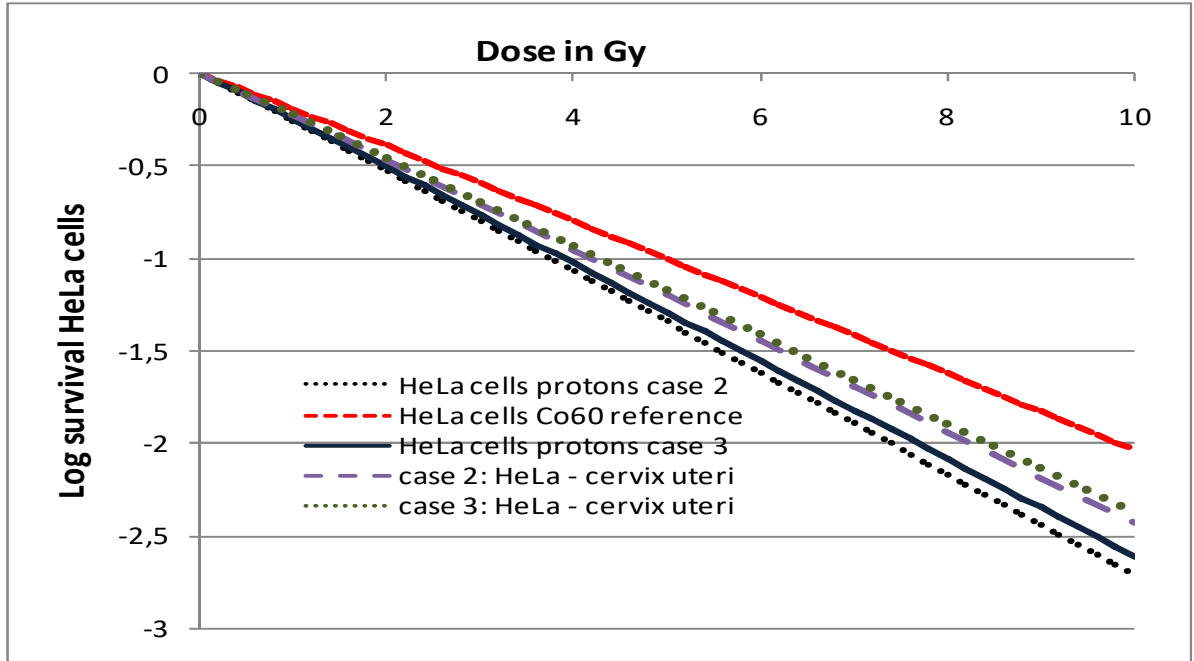

**Figure 12:** Survival curves of HeLa cells according to Figure 7 and Figure 11.

**Table 3:** Parameters of the HeLa cells and cervical ueteri.

| HeLa cells | approach 2 | A | a | b | approach 3 | A | a | b |
|---|---|---|---|---|---|---|---|---|
| | | 1.07 | 0.634 | 0.305 | | 1.07 | 0.61 | 0.293 |
| Cervical uteri | approach 2 | A | a | b | approach 3 | A | a | b |
| | | 1.07 | 0.567 | 0.272 | | 1.07 | 0.55 | 0.265 |

# 4. Discussion

It should be mentioned that there have been attempts to determine LET and RBE and the connection between properties of protons and γ-rays in a rather different way[15,16]. However, it appears that these authors do not provide a method to calculate this connection in a stringent way, in particular, with regard to the connection between ionization density of high LET-radiation and low LET as realized by $Co^{60}$.

The present analysis of LET and RBE is based on some simplified assumption, which may, in general, not be sufficient: 1. The 3D-problem is restricted to identical pencil beams by shifts to the x/y-plane in order to form clinically realistic target volumes, i.e. the substitution (14) has to be accounted for. 2. A further simplifications refers to eqs. (31a, 31b): Thus it is assumed that the necessary dimension length is in both cases identified with $l_c$ and the amplitude parameter A is identical when passing from reference system to any other system. In the 3D-case the parameter $l_c$ might be different in the x/y-directions. In the present study only λ = a + b is scaled to receive a different RBE-value. 3. The present restriction does not account for the lateral scatter of proton beams. In a formal description it is possible to treat this aspect by deconvolutions. However, in clinical applications this procedure might be intricate, and the extension to the 3D-procedure with different values for $α_x$, $α_y$, and $α_z$ cannot be avoided, in particular, if different beam directions are applied. 4. The debate on the correct clinical RBE for protons should not assume a dogmatic character, since only under conditions realized in cultures or spheroids a rigorous fixation of the S-value is possible. Moreover, the RBE might slightly be depend on the beam-guide, although this effect is small as verified in Figure 12.

It should be mentioned that there have been attempts to determine LET and RBE and the connection between properties of protons and γ-rays in a rather different way[20,21]. However, it appears that these authors do not provide a method to calculate this connection in a stringent way, in particular, with regard to the connection between ionization density of high LET-radiation and low LET as realized by $Co^{60}$.

Finally we should add that the subtraction method for the LET determination we have used in this study can also be extended to heavy carbons in order to reach some information about RBE of radiation with increased ionization density based on spatial diffusion distribution of ionized biomolecule

## Appendix: General solution method of the spatial term H

In the preceding section we have used the constraint that H(z) (or in the 3D case H(x, y, z) to a solitary solution is sufficient for the present investigation. This might, however, be a rather strong restriction. The general solution spectrum is obtained a more general procedure.

We should recall that the solitary solution (12) has the advantage to handle in an easy fashion, but they do not represent the complete solution spectrum of eq. (11), which is provided by the expansion:

$$H_k(z) = \sum_{k=n}^{\infty} c_n \cdot \sec h^n(\alpha \cdot z) \quad (34)$$

Thus n is running from 1 to M (M → ∞), and for each starting index (n = 1, 2,...,M) the expansion of a solution function is provided. With respect to the calculation procedure of the coefficients $c_n$ we make use of a previous elaboration[6] yielding the sequence $c_1 \to c_2 \to c_3 \to c_4$ etc. with $c_1$ as the only free coefficient to be determined by the normalization, i.e. $c_n = c_n(c_1)$. Thus we only consider the most interesting case k = 1, and n is running from 1 to M (M →∞). The following abbreviations are valid:

$$\lambda_1 = B \cdot \lambda \,;\, \lambda_2 = \lambda \cdot A;\, \lambda = a + b;\, u = -\lambda_2 / \lambda_1 \quad (35)$$

Then we obtain:

$$c_1 \cdot (\alpha_0 \cdot \alpha^2 - \lambda_1) = 0 \quad (36)$$

The expansion up to order n = 3 is given by:

$$\left. \begin{array}{l} c_2 = -\lambda_2 \cdot c_1^2 /(3 \cdot \lambda_1) \\ c_3 = c_1/4 + \lambda_2^2 \cdot c_1^3 /(12 \cdot \lambda_1^2) \end{array} \right\} (37)$$

The general formation law of $c_n$ reads:

$$\left. \begin{array}{l} c_n = n \cdot c_1^n \cdot u^{n-1} / 6^{n-1} \\ + \sum_{r=1}^{M} f_r(n) \cdot c_1^{n-2r} \cdot u^{n-2r-1} /(3^{n-1-2r} \cdot 2^{n-1}) \\ f_r(n) = \dfrac{(n-2r)^2}{r!} \cdot \prod_{j=1}^{r-1}(n-j) \end{array} \right\} (38)$$

The expansion (34) is characterized by alternating signs; $c_n$ (n: odd) is always positive, whereas $c_n$ (n: even) is throughout negative. Thus $c_n$ (n→∞) vanishes and the Leibniz convergence criterion is applicable. In practical problems it is sufficient to account for terms up to order 4.

In the 3D case the following substitution is also applicable:

$$\mathrm{sech}^n(\alpha \cdot z) \rightarrow \mathrm{sech}^n(\vec{\alpha} \cdot \vec{z}) \quad (39)$$

In eq, (39) we now have to put $\alpha^2 = \alpha_x^2 + \alpha_y^2 + \alpha_z^2$ as previously carried out. If we would start with a different kind of expansion according to eq. (34) we obtain eq. (39) in the form:

$$\left.\begin{array}{l} c_1 \cdot (\alpha_0 \cdot \alpha^2 \cdot k^2 - \lambda_1) = 0) \\ k = 2, 3, \ldots, M \end{array}\right\} \quad (40)$$

These modifications yield different patterns of the diffusion behavior.

Since the parameters $a$ and $b$ ($\lambda = a + b$) stand in very close relationship to $\alpha_0 \cdot \alpha^2$, we regard eqs. (17) and (21). Thus for k = 1 the correspondence is reduced to eq. (17)., which is associated with the broadest ion concentration: $\rho = \alpha_0 \cdot \alpha^2 / B$, whereas for k > 1 the density of the deposited energy increases proportional to $k^2$. By that, $\lambda$ is growing in the same manner, and increasing a and b imply the behavior of the survival function, which tells us that the shoulder becomes throughout smaller and smaller and the steepness of S significantly increases. With regard to the determination of $\lambda \cdot A$ the expansion (19) is difficult to handle, since the term $\lambda_2 = \lambda \cdot A$ is connected with the coefficients $c_n$ and appear as a power expansion. Therefore it is more convenient to restrict the determination of A to the homogenous case according to eq. (13a), which, however, only represents a solitary solution.

In the case of the expansion coefficients $c_n$ according to eq. (34) and by taking account for the substitutions (19) the resulting equation for D = 0 and z = 0 eq. (25) would read:

$$S = 1 \rightarrow A \cdot \sum_{n=1}^{\infty} c_n / (1 + B) = 1 \quad (41)$$

In practical applications it is reasonable to restrict eqs. (34, 40) to a finite value, e.g. M = 3.

The question arises, whether generalizations such as by eq. (34) may be useful in detector problems. The only application we at present can verify is the carrier diffusion in scintillators with relevance of diffusion band structures[21].

## Conflict of interest
The author declares there is no conflict of interest. The author alone is responsible for the content and writing of the paper.

## .References